\documentclass[12pt]{article}
\usepackage{epsf,latexsym,psfig}
\usepackage{amsfonts,amssymb} 
\epsfverbosetrue
\newcommand{\double}[1]{\mathbb{#1}}

\newcommand{\rr}{\double{R}}

\newcommand{\de}{\hbox{\rm{d}}}

\newcommand{\bb}{\begin{eqnarray}}
\newcommand{\ee}{\end{eqnarray}}
\newcommand{\eee}{\nonumber\end{eqnarray}}
\newcommand{\qq}{\quad}

\begin{document}

\font\twelve=cmbx10 at 13pt
\font\eightrm=cmr8

\thispagestyle{empty}

\begin{center}
${}$
\vspace{3cm}

{\Large\textbf{From Hubble diagrams to scale factors}} \\

\vspace{2cm}

{\large
Thomas Sch\"ucker\footnote{at Universit\'e de Provence,
schucker@cpt.univ-mrs.fr } (CPT\footnote{Centre de Physique
Th\'eorique\\\indent${}$\qq\qq CNRS--Luminy, Case
907\\\indent${}$\qq\qq F-13288 Marseille Cedex 9\\\indent${}$\qq
Unit\'e Mixte de Recherche (UMR 6207)
du CNRS et des Universit\'es Aix--Marseille 1 et 2\\
\indent${}$\qq et Sud
Toulon--Var, Laboratoire affili\'e \`a la FRUMAM (FR 2291)}),
Andr\'e Tilquin\footnote{tilquin@cppm.in2p3.fr }
(CPPM\footnote{Centre de Physique des Particules de
Marseille\\\indent${}$\qq\qq CNRS--Luminy, Case
907\\\indent${}$\qq\qq F-13288 Marseille Cedex 9\\\indent${}$\qq
Unit\'e Mixte de Recherche (UMR 6550)
du CNRS et de l'Universit\'e Aix--Marseille 2}) }

\vspace{3cm}

{\large\textbf{Abstract}}
\end{center}
We present a lower bound on the radius of the universe today $a_0$ and a
monotonicity constraint on the Hubble diagram. Our theoretical input is
Einstein's kinematics and maximally symmetric universes. Present
supernova data yield $a_0 > 1.2\cdot
10^{26}$ m. A first attempt to quantify the monotonicity constraint is
described. We do not see any indication of non-monotonicity.

\vspace{2cm}

\noindent PACS: 98.80.Es, 98.80.Cq\\
Key-Words: cosmological parameters -- supernovae
\vskip 1truecm

\noindent CPT-2005/P031\\
\noindent astro-ph/05mmxxx
\vspace{2cm}

\section{Introduction}

In a homogeneous, isotropic, expanding universe, the
apparent luminosity of a standard candle is a monotonically decreasing
function of the time of flight of emitted photons if the universe is open.
This is also true in spherical universes if the time of flight is small
enough with respect to the radius divided by the speed of light.
In principle the (apparent) luminosity $\ell(t)$ as a function of time can
be used to measure the scale factor $a(t)$. In reality, arriving photons do
not tell us their time of flight, but only their spectral deformation,
$z:=\nu _{\rm table}/\nu _{\rm observed}-1$. In an expanding universe
$z$ is positive, `red shift'. If we pretend to know the scale factor $a(t)$ we
can compute the luminosity
$\ell(z)$ and confront it to the Hubble diagram. The transform
$a(t)\rightarrow \ell(z)$ reminds us of the Fourier transform and of
course we are interested in the inverse transform $\ell(z)\rightarrow
a(t)$. Therefore we must ask three questions: What is the domain of
definition of the initial transform, what is its
image and is the transform injective? Should the
measured luminosity $\ell(z)$ be `far away' from the image our working
hypotheses are put to test.

\section{The hypotheses}

We assume the kinematics of general relativity: $(i)$ The gravitational
field is coded in a time-space metric of signature $+---$, the configuration
space is the set of all such metrics. $(ii)$ Massive and massless, pointlike
test particles,
subject only to gravity, follow timelike and lightlike geodesics.
$(iii)$ Pointlike clocks, e.g. atomic clocks, are necessarily massive. They
move on timelike curves (not necessarily geodesics) and indicate proper
time
$\tau $.

This kinematics is covariant under general coordinate transformations.
In 1983 the meter was officially abandoned as fundamental unit in favor
of an absolute speed of light and we would like to stress that at least since
then any non-covariant kinematics is void of any physical meaning.

We also assume the hypotheses of spatially maximally symmetric
cosmology: $(iv)$ The metric is Robertson-Walker,
\bb c^2\de \tau ^2=c^2\de t^2-a(t)^2\left[ \de\vec x^2+k\,\frac{(\vec
x\cdot \de\vec x)^2}{1-k\vec x^2}\, \right] .\ee
The scale factor $a(t)$ is a strictly positive function of time, $k=-1,\ 0,\ 1$
for the pseudosphere, the Euclidean space and the sphere. We take the
coordinates $\vec x$ dimensionless and call them `co-moving position',
while the scale factor is measured in meters. For closed universes, $k=1$,
we must restrict the spatial coordinates to the unit ball,
$|\vec x|<1$, they describe the northern hemisphere. For open universes,
$\vec x$ varies in $\rr^3$. $(v)$ The
test particles are (superclusters of) galaxies and photons. The former
follow the timelike geodesics $t=\tau ,\
\vec x={\rm constant}$. Note that due to the high degree of symmetry the
proper time is universal for all these timelike geodesics and is taken as
time coordinate. The photons are emitted from the a galaxy at time $t$ and
arrive at our position at time
$t_0$, today.

The symmetry hypothesis and the choice of the coordinates $(t,\vec x)$
have reduced the configuration space to the set of positive functions
$a(t)$ on the real line and to the para\-meter $k=0,\ \pm 1$.

\section{The Hubble diagram}

>From these hypotheses we can compute \cite{berry} the (apparent)
luminosity
$\ell(t)$ of a standard candle in Watt$/{\rm m}^2$ as a function of
emission time $t$
\bb \ell(t)= \,\frac{L}{4 \pi a^2_0}\, \,\frac{a(t)^2}{a^2_0\,s^2(\chi
(t))}\, \ee
where $L$ is the absolute luminosity of the standard candle in Watt,
$a_0:=a(t_0)$ is the scale factor today,
\bb \chi(t):=\int_t^{t_0}\,\frac{c\,\de\tilde t}{a(\tilde t)}\, \ee
is the dimensionless co-moving geodesic distance covered by the photon
and
\bb s(\chi ):=\left\{ \matrix{\sin \chi ,& k=1\cr \chi ,& k=0\cr \sinh
\chi ,& k=-1}\right. .\ee
The luminosity has a singularity at $t=t_0$, ` short distance divergency',
for any value of $k$ and any scale factor. For closed universes, $k=1$ we
may have additional singularities for $\chi =\pi $, `antipode
divergencies', for $\chi =2\pi $ we are back at a second short distance
divergency and so forth if $\chi $ continues to grow. Of course there
might be a horizon, i.e. an upper bound for
$\chi $, masking all or some of the singularities.

>From the above hypotheses we can also compute the spectral
deformation as function of emission time,
\bb z(t)=\,\frac{a_0}{a(t)}\, -\,1.\ee
The theoretical Hubble diagram is the parametric plot in the $z-\ell$
plane as $t$ varies.

If we suppose that the scale factor is strictly
increasing with $\dot a:=\de a/\de t>0$ then $z$ is positive, `red shift', and
we can invert the function
$z(t)$. By abuse of notations we write $t(z)$ for its inverse. Then the
Hubble diagram is a function $\ell(t(z))$ that still by abuse is written
$\ell(z)$,
\bb\ell(z)=\,\frac{L}{4 \pi a^2_0}\, \,\frac{1}{(z+1)^2\,s^2(\chi
(z))}\, \ee
where
\bb \chi(z):=\chi (t(z))=\,\frac{c}{a_0}\, \int_0^{z}\,\frac{\de\tilde
z}{H(\tilde z)}\,\ee
and $H(z):=\dot a(t(z))/a(t(z))$ is the Hubble rate.

The short distance divergency now is at $z=0$ and easy to get rid of: let us
define the regularized luminosity
\bb f(z):=z^2\ell(z)=\,\frac{L}{4 \pi a^2_0}\, \,
\frac{z^2}{(z+1)^2\,s^2(\chi (z))}\, .\label{reg}\ee
Indeed,
\bb
f(0)=\,\frac{L}{4\pi c^2}\,H_0^2 ,\qq f'(0)=
\,\frac{L}{4\pi c^2 }\,H_0^2\,(q_0-1) ,\label{fzero}\ee
where the prime is differentiating with respect to $z$, $q(z)$ is the {\it
deceleration} parameter,
\bb q(z)=-\,\frac{a\ddot a}{\dot a \dot a}(t(z)),\ee
and $H_0:=H(0),$ $q_0=q(0)$.

\section{The transform $a(t)\longrightarrow \ell(z)$}

Let us first try to describe the image, that is all luminosity functions
$\ell(z)$ which can be obtained from strictly increasing scale factors
$a(t)$ with $\dot a>0$ and with $k=0,\ \pm 1$.

We already know that
$\ell(z)$ comes with the short distance divergency at
$z=0$ which is such that $z^2\ell(z)=:f(z)$ is regular there.
For open universes there are no other singularities. Indeed,
$(z+1)^2\ell(z)=:g(z)$ is a decreasing function. For closed universes on
the other hand,
$g(z)$ goes through a minimum as the photons pass the equator, $\chi
=\pi /2$. From there on the luminosity increases again and goes to the
antipode divergency. It might of course happen that the equator is
masked by the horizon in which case $g(z)$ remains decreasing for ever,
even though the universe is closed.

Let us now ask whether the transform is injective in the domain of
increasing scale factors.

If we pretend to know $k$ the answer is
affirmative for open universes. Indeed, solving equation (\ref{reg}) for
$s(\chi )$ and differentiating with respect to $z$ yields \cite{gef}
\bb \,\frac{a_0}{c}\, H(z)=\,\frac{(z+1)\sqrt{4\pi a_0^2
L^{-1}(z+1)^2f(z)-kz^2}}{1-{\textstyle\frac{1}{2}} z(z+1)f'(z)/f(z)}\,=
\,\frac{z(z+1)\sqrt{4\pi a_0^2
L^{-1}g(z)-k}}{1-{\textstyle\frac{1}{2}} z(z+1)f'(z)/f(z)}\, .\ee
Therefore we can reconstruct the Hubble rate from the luminosity.

Integrating the Hubble rate with respect to $t$ gives us the
scale factor with the ambiguity of the initial condition $a_0$. But for flat
universes this initial condition is unphysical, by
a coordinate transformation of $\vec x$, we can set $a_0=1$ m. This is
different for curved universes where $a_0$ is related to a local
observable, curvature. In the closed case, $a_0$ is also related to a global
observable, the
radius of the universe today. However, unless $g(z)$ already exhibits an
increase, only a lower bound on the radius today can be reconstructed
from the luminosity,
\bb a_0\ge \sqrt{\,\frac{L}{4\pi g_{\rm min}}\, },\qq g_{\rm min}:=
\min_{z>0}\{(z+1)^2\ell(z)\}.\label{ineq}\ee
Note that this lower bound does not depend on the absolute luminosity $L$.

If we admit that we do not know $k$ and if the luminosity function
$\ell(z)$ satisfies: $(i)$ $z^2\ell(z)$ is regular at $z=0$, $(ii)$
$(z+1)^2\ell(z)$ is decreasing, then there are three positive functions
$a_\pm(t)$ and
$a(t)$, such that the universes with scale
factor
$a_-(t)$, $k=-1$, with scale factor $a(t)$, $k=0$ and with scale
factors $a_+(t)$, $k=1$ have the same luminosity function $\ell(z)$. These
three scale factors satisfy
\bb a_-(t_0)\,\sinh\chi _-=a(t_0)\,\chi =a_+(t_0)\,\sin \chi
_+.\label{three}\ee
Note that in the flat case the `initial condition'
$a(t_0)$ carries no information while in the closed case $a_+(t_0)$ must
satisfy the inequality (\ref{ineq}).
\vskip.3cm\noindent
{\bf Example} (constant deceleration parameter):\\
Take the scale factor,
\bb a(t):= a_0(pH_0t)^{1/p},\qq p>0,\qq
0<t<t_0=\,\frac{1}{pH_0}\,.\label{scalex1}\ee
Then the Hubble rate is
$H=H_0(z+1)^p$ and the deceleration parameter is constant, $q\equiv
p-1$. The co-moving distance is
\bb\chi =\,\frac{c}{a_0H_0}\, \ln
(z+1)=-\,\frac{c}{a_0H_0}\,\ln (H_0t),\qq\qq p=1,\ee
\bb\chi =
\,\frac{c}{(p-1)a_0H_0}\, (1-(z+1)^{1-p})=\,\frac{c}{(p-1)a_0H_0}\,
(1-(pH_0t)^{1-1/p}),\qq\qq p\not=1.\ee
For $p>1$, there is a horizon at
$\chi =c/((p-1)a_0H_0)$. In particular for $p=2$, $k=0$
the regularized luminosity $f(z)$ is constant.
\vskip.3cm\noindent
{\bf Example} (constant regularized luminosity):\\
Suppose we have measured a constant regularized luminosity $f(z)\equiv
LH_0^2/(4\pi c^2)$. Then the Hubble rate is
\bb H=(z+1)\sqrt{H_0^2(z+1)^2-kc^2z^2/a_0^2}.\ee
The three solutions of these three differential equations, $k=\pm 1,\ 0$,
in terms of the scale factors
are obtained from equations (\ref{three}) and (\ref{scalex1}):
\bb a(t)= a_0(2H_0t)^{1/2} \sqrt{1-k\left[
\,\frac{1-(2H_0t)^{1/2}}{a_0H_0/c}\,\right] ^2},\qq
0<t<t_0=\,\frac{1}{2H_0}\,.\ee
To alleviate notations we have suppressed the subscripts $\cdot_\pm$
from $a(t)$ and $a_0$. In the closed case,
$a_0$ must satisfy the inequality (\ref{ineq}):
\bb a_0\ge\,\frac{c}{H_0} \,\frac{z_{\rm max}}{z_{\rm max}+1}\, \ee
and the inflection point of $g(z)$ is hidden behind the horizon at $\chi
=\arcsin [c/(a_0H_0)]$.
\vskip.3cm\noindent
{\bf Example} (constant Hubble rate):\\
To have an example without horizon consider the scale factor
\bb a(t)= a_0\exp[H_0(t-t_0)].\ee
This is a limiting case of the first example with $p\rightarrow 1$.
It has constant Hubble rate and constant deceleration parameter,
$H(z)\equiv H_0$,
$q(z)\equiv -1$, and $\chi =cz/(a_0H_0)$. In the closed case, $g(z)$ has an
infinite number of inflection points alternating with short distance
divergencies. The inflection points are all minima,
the first is located at $z_{\rm infl}=\pi a_0H_0/(2c)$.
\vskip.3cm\noindent
{\bf Example} (closed universe):\\
Suppose we have measured the luminosity:
\bb \ell(z)=\,\frac{LH_0^2}{4\pi c^2} \,\frac{z+2}{z^2}.\ee
It is strictly decreasing and has the
correct short distance divergency.
Its $g(z)$ has one and only one inflection point at $z_{\rm
infl}=(1+\sqrt{17})/2\sim\,2.56$ and suggests a closed universe with
$a_0\sim 0.337 c/H_0$.
\vskip.3cm\noindent
{\bf Counter-example} (wiggling $g(z)$):\\
Suppose we have measured the luminosity, figure 1:
\bb \ell(z)=\,\frac{LH_0^2}{4\pi c^2} \,\frac{(\sin z/z)^2+0.1}{z^2}.
\label{ellw}\ee
Again it is strictly decreasing and has the
correct short distance divergency.
Now $g(z)$ has a maximum, figure 2. Therefore
no Robertson-Walker universe, neither open nor closed, exists with this
luminosity.

\begin{figure}[h]
\epsfxsize=8cm
\hspace{3.6cm}
\epsfbox{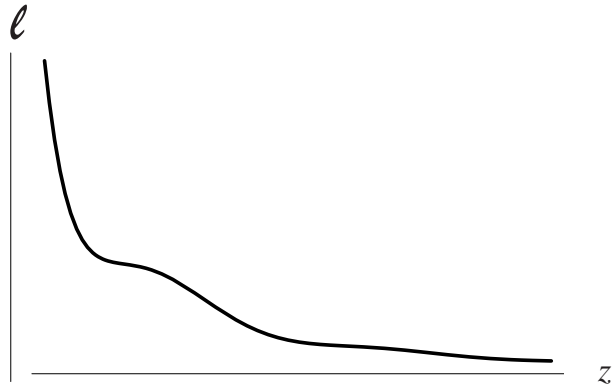}
\caption{The monotonic luminosity (\ref{ellw})}
\end{figure}

\begin{figure}[h]
\epsfxsize=8cm
\hspace{3.6cm}
\epsfbox{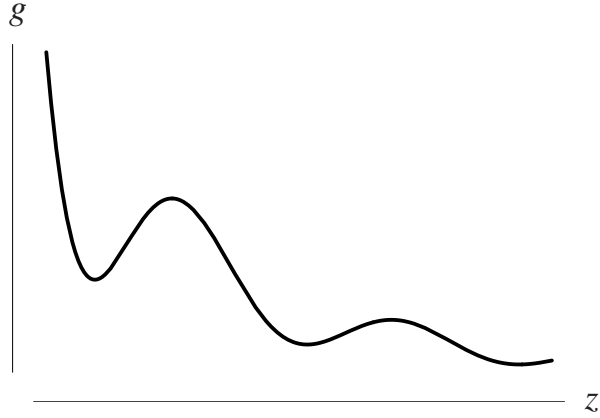}
\caption{Its wiggling $g(z)$}
\end{figure}

The last two examples illustrate that our constraint of monotonic $g(z)$
is stronger than the constraint of monotonic luminosity $\ell (z)$.

\section{Non-monotonic scale factors}

If the scale factor is strictly decreasing we get similar results with a
negative spectral deformation: $-1<z<0$, `blue shift'.

One might think that one can produce non-monotonic functions $g(z)$
by starting from non-monotonic scale factors $a(t)$. This is not true. In
fact any non-monotonic scale factor produces multivalued luminosities
in terms of the spectral deformation $z$. The first example is of course
the constant scale factor with no spectral deformation, $z\equiv 0$, but
varying luminosity. A more generic example is, see figure 3:
\begin{figure}[h]
\epsfxsize=7cm
\hspace{4.1cm}
\epsfbox{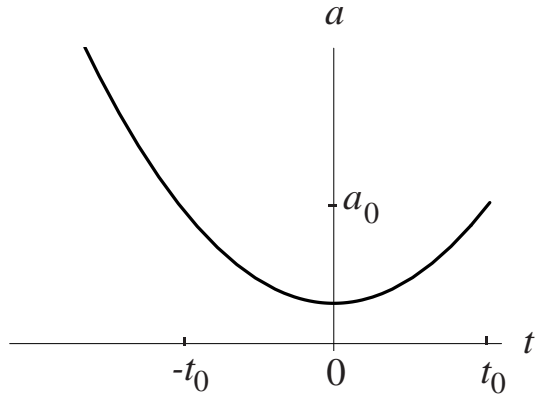}
\caption{The non-monotonic scale factor (\ref{non-m})}
\end{figure}
\bb a(t)= {\textstyle\frac{1}{2}} gt^2+\alpha ,\qq g,\alpha
>0.\label{non-m}\ee
For positive $t_0$, we have a maximal redshift of
$z_{\rm max}=a_0/\alpha -1$. We have $z=0$ for $t=\pm t_0$ and a blue
shift for $t<-t_0$:
\bb-1<z_{\rm min}=\,\frac{a_0}{{\textstyle\frac{1}{2}} gt^2+\alpha }\,
-1<0.\ee
The dimensionless co-moving distance is
\bb \chi (t) =c\sqrt{\,\frac{2}{g\alpha }\, }\left[ \arctan\left(
\sqrt{\,\frac{g}{2\alpha }\,} t_0\right) -\arctan\left(
\sqrt{\,\frac{g}{2\alpha }\,} t\right)\right]\ee
and the relation between emission time $t$ and spectral deformation $z$
is
\bb t(z)=\pm \sqrt{\,\frac{2}{g}\, }\sqrt{\,\frac{a_0}{z+1}\, -\alpha }.\ee
Note that the Hubble rate vanishes at the inflection time $t=0,\
z=z_{\rm max}$ while the deceleration parameter diverges there. Note
also that the regularized luminosity vanishes at $t=-t_0,\ z=0$. As $t$
tends to minus infinity
$z$ tends to
$-1$ and the luminosity, regularized or not, diverges, the
`ultra-violet divergency'. In the closed case, we have in addition antipode
and short distance divergencies. Figure 4 shows only one antipode
singularity.

\begin{figure}[h]
\epsfxsize=12cm
\hspace{2.0cm}
\epsfbox{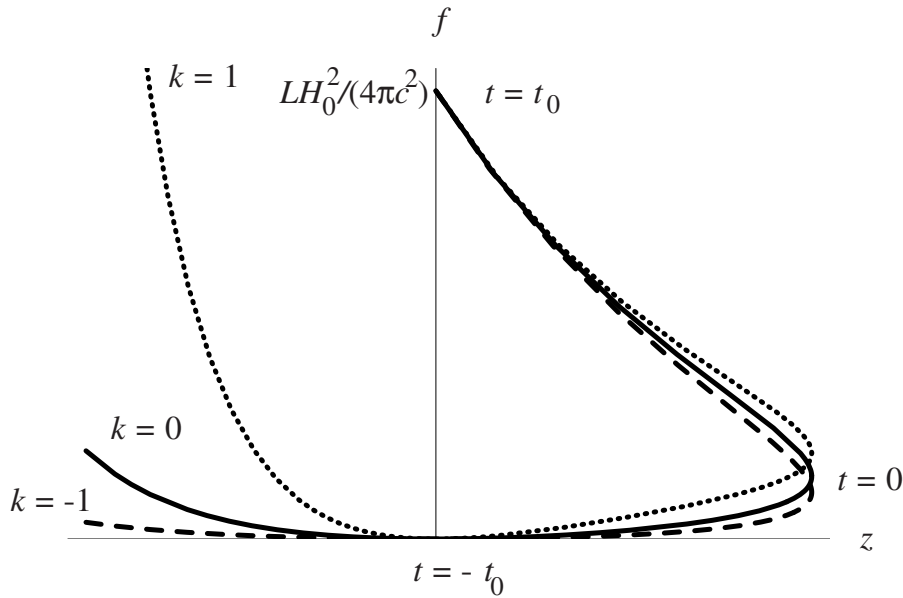}
\caption{The regularized luminosity of the non-monotonic scale factor
(\ref{non-m})}
\end{figure}

\section{Data and conclusion}

We use the 'Gold' sample data compiled by Riess et al. (2004) \cite{r}, with
157 SN's including a few at $ z>1.3$ from the Hubble Space Telescope (HST
GOODS ACS Treasury survey). For convenience we normalize the
luminosity to the maximum absolute SN luminosity estimated by Jha et al.
(1999) \cite{jj}, Saha et al.(2001) \cite{ss} and Gibson \& Stetson (2001)
\cite{gs},
$L=L_{max}=(1\pm 0.1)\cdot 10^{35}$ W. The Hubble rate today $H_0$ is
taken as
$(70\pm 5)$ km ${\rm s^{-1}\ Mpc^{-1}}$ \cite{kr,dd}.

\begin{figure}[h]
\epsfxsize=10cm
\hspace{2.5cm}
\epsfbox{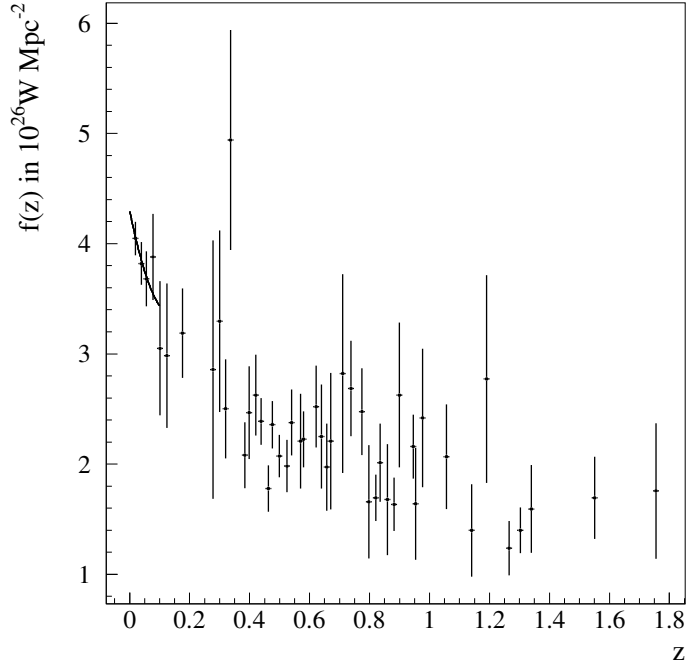}
\caption{The regularized luminosity as measured today \cite{r}, with a
binning of 0.02 in red shift. The full line
at low red shift corresponds to a second order polynomial
extrapolation fit.}
\end{figure}

The regularized luminosity allows to extract
$LH_0^2$ from the Hubble diagram with small red shift, figure 5.
The value $f(0)$ is extracted by a second order polynomial extrapolation
fit on the SN data up to a red shift of 0.1. By construction, the fitted
value $f(0)$ is equal to
$LH_0^2/(4\pi c^2) = (4.3\pm 0.4)\cdot 10^{26}\ {\rm W\ Mpc^{-2}}$ where
the error is only coming from the fit itself.

\subsection{Lower bound on the radius of the universe today}

The first of equations (\ref{fzero}) and equation (\ref{ineq}) give the
lower bound on
$a_0$ as
\bb a_0\ge \sqrt{\,\frac{f(0)}{g_{\rm min}}\, } \,\frac{c}{H_0}\,. \ee
The minimal value of
$g(z)$ is obtained from the SN recorded at $z=1.3$ and is equal to $g_{\rm
min} = (3.83 \pm 0.72)\cdot 10^{26}\ {\rm W\ Mpc^{-2}}$. At $95\%$
confidence level (CL) we have the upper bound on the minimal value of
$g(z)$:
$ g_{\rm min}<5.6\cdot 10^{26}\ W\ Mpc^{-2}$. Combining the errors from
$f(0)$ and $g_{\rm min}$ gives the
lower bound on $a_0$ at $95\%$ CL:
\bb a_0 > 0.88\ c/H_0\sim 3.8\ {\rm Gpc }\ \sim\ 1.2 \cdot 10^{26}\ {\rm
m}\
\sim 1.3\cdot 10^{10}\ {\rm light\ years}.\ee
This lower bound on $a_0$ translates into a lower bound on the
curvature density,
\bb\Omega_k:=-k\left( \,\frac{c}{a_0 H_0}\,\right) ^2 \ >\ -1.29\qq{\rm
at
\ 95\%\ CL.}\ee
Note that this limit is independent of the values of the
absolute luminosity
$L$ and of the Hubble rate today.
It is also independent of any dynamical hypothesis.

Let us compare this bound with the one obtained from the SN data but
now adding the dynamics of the $\Lambda CDM$ cosmology fitting the
matter density $\Omega _m$, the cosmological constant density $\Omega
_\Lambda $ and the nuisance normalisation parameter and without any
other input constraint: $\Omega _k> -1$ at 95 \% CL.

\subsection{Wiggles and non-monotonicity}

We must now ask the question whether the data is compatible with a
monotonic luminosity $\ell(z)$. We will also ask the finer question
whether $g(z)$ is monotonic.

To detect non-monotonicity in the SN data set we assume that the
luminosity $\ell(z)$ and its $g(z)$ can be described by monotonic
functions to which we add a simple Gaussian:
\bb L(z) = \ell(z,LH_0^2,p) + a_w \exp{-{(z-z_w)^2}/{(2 \delta z_w^2)}}
\label{mod1}\ee
and
\bb G(z) = g(z,LH_0^2,p) + a_w \exp{-{(z-z_w)^2}/{(2 \delta z_w^2)}}.
\label{mod2}\ee

The monotonic functions $\ell(z,LH_0^2,p)$ and $g(z,LH_0^2,p)$ are
derived from a power law parameterization of the scale factor $a(t):=
a_0(pH_0t)^{1/p}$ (constant deceleration parameter),
$1/5<p<1$, $k=0$. It yields the monotonic luminosity
\bb \ell(z,LH_0^2,p)=\,\frac{LH_0^2}{4\pi c^2}\,\left(
\frac{p-1}{(z+1)[1-(z+1)^{1-p}]}\right) ^2\, ,\ee
and with $p=0.69$ it
describes well the
$\Lambda CDM$ cosmology up to a redshift of 1.8 \cite{ff}. Its
acceleration is positive with $q\equiv -0.31$.

Our wiggle detection
procedure
consists of scanning the plane in wiggle position
$z_w$ and wiggle width $\delta z_w$ in steps of 0.01 in both directions.
In each point $(z_w,\delta z_w)$ of this plane we
fit the normalization $LH_0^2$, the power $p$
and the wiggle amplitude $a_w$.

Warning: if the wiggle amplitude $a_w$ is smaller than a critical
amplitude
$a_w^c$ the modified functions (\ref{mod1}) and (\ref{mod2}) will still
be monotonic.

We would claim that $\ell(z)$ and a fortiori the luminosity $g(z)$ is
not monotonic if the ratio between the fitted wiggle
amplitude and the associated error is greater than 5 (5$\sigma$ level
detection) and if the wiggle amplitude is greater than the critical one.
The sensitivity of the method is computed by Monte Carlo simulation. The
same SN sample than the Riess data set with the same statistical power is
simulated assuming the
$\Lambda\rm{CDM}$ cosmology and a wiggle of positive or
negative amplitude is added to $\ell(z)$ and $g(z)$ for each
point in the ($z_w$,$\delta z_w$) plane. We apply the wiggle detection
procedure on each simulation, restricted to a small grid of points
around the simulated one to speed up the processing. The significance on
the wiggle amplitude
$(|a_{w\,\rm{fitted}})|/\sigma_{a_w})$ is computed in each point and the
smaller value from positive or negative wiggle amplitude is retained. The
sensitivity is computed at a $2\sigma$ level corresponding to a $95\%$
confidence level exclusion limit on the wiggle magnitude $m_w$ defined
by
\bb a_w(z_w) = \pm(10^{-m_w/2.5}-1)\ell(z_w).\ee

Figure 6 shows the significance of the wiggle fit
performed on the Riess data sample (color contours) for the luminosity
$\ell(z)$ and
$g(z)$ with $z_w$ varying from 0.01 to 1.8 and $\delta z_w$
from 0.01 to 2 in steps of 0.01 in both directions. The maximum
significance for both $\ell(z)$ and $g(z)$ is 2.4 for a wiggle at the position
$z_w= 0.45$, with width of 0.07. The wiggle magnitude is $m_w= 0.16$ for
the luminosity $\ell(z)$ and $m_w=0.25$ for $g(z)$. The dashed lines
indicate the location of the critical wiggle magnitude $m_w^c$ of a
positive wiggle that breaks the monotonicity.

No wiggle greater than this value is observed and we conclude that no
wiggle is detected at a $5\sigma$ level using the actual SN data set. On
the same figures, the sensitivity for different values of the wiggle
magnitude is shown (plain line). Wiggles of magnitude greater than
$2$ are excluded at $95\%$ CL up to a redshift of 1.6. Up to a redshift of 1,
the $95\%$ CL exclusion limit on the wiggle magnitude is
$ 0.6$. These two magnitudes are below the critical magnitudes and
therefore these wiggles do not upset the monotonicity.

\vfil\eject
\thispagestyle{empty}

\begin{figure}[h]
\hspace{1,6 cm}
\begin{minipage}[t]{6 cm}
\vspace{-2 cm}
\epsfxsize=12cm
\epsfbox{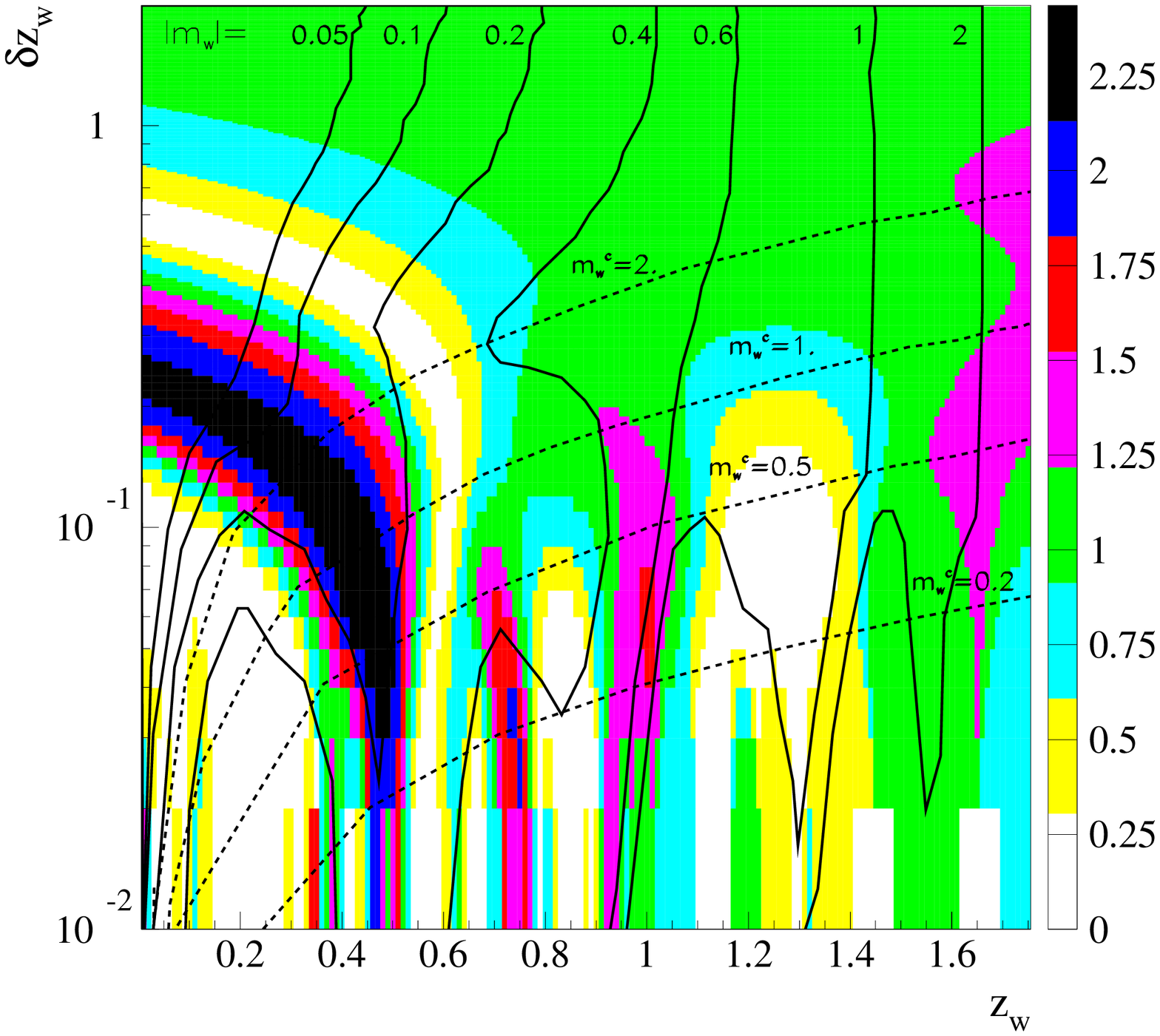}
\epsfxsize=12cm
\epsfbox{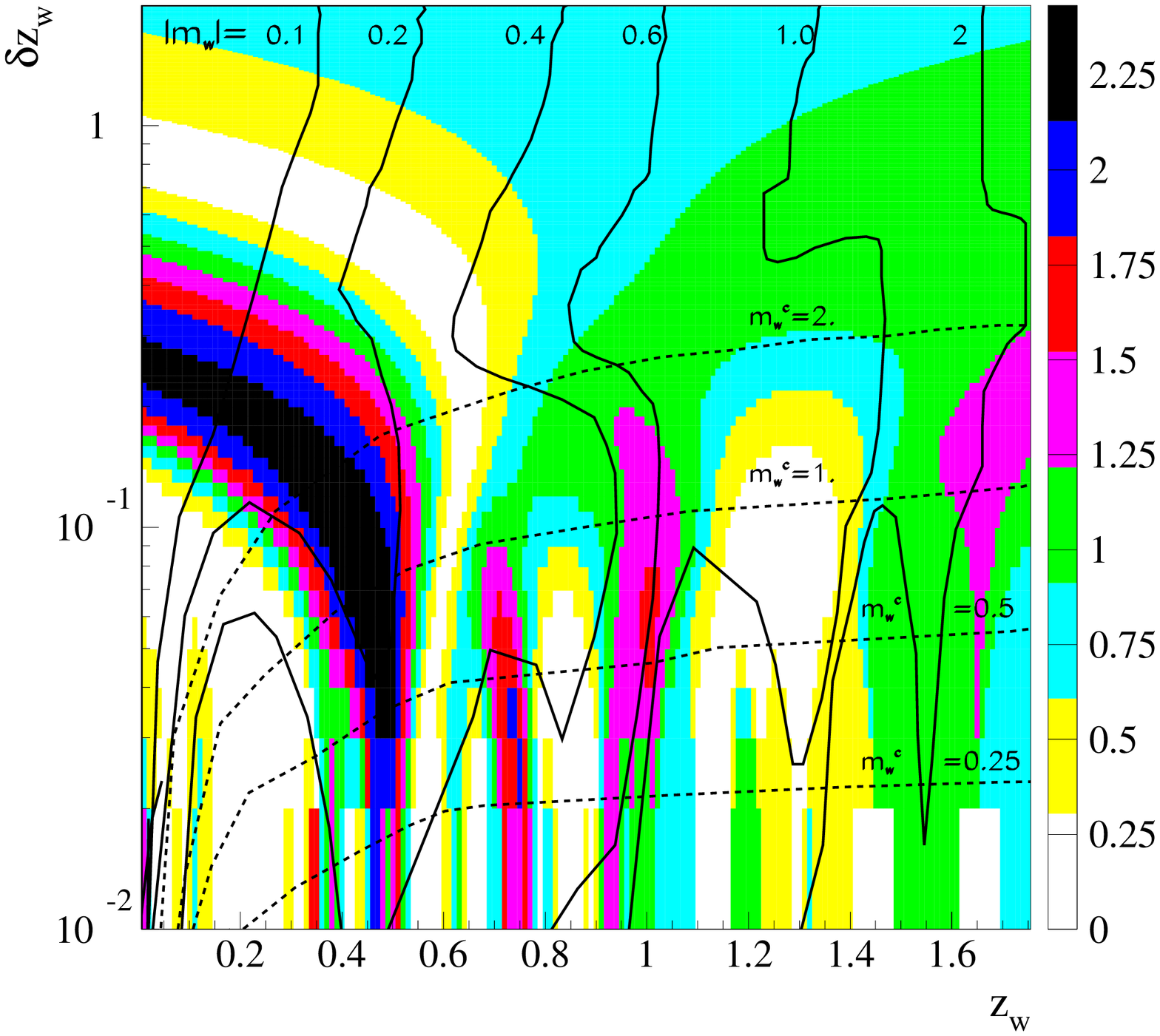}
\end{minipage}
\caption{Color contours: Significance of wiggle detection (vertical
colour scale) as function of the
wiggle position $z_w$ and width $\delta z_w$, logarithmic scale. Full
lines: Expected
$2\sigma$ sensitivity of the wiggle detection as a function of wiggle
magnitude $m_w$. The dashed lines
indicate the location of the critical wiggle magnitude $m_w^c$ of a
positive wiggle that breaks the monotonicity. The upper panel is the
exclusion plot for a
non-monotonic luminosity
$\ell(z)$, the lower panel for non-monotonic $g(z)$.}
\end{figure}


\begin{thebibliography}{10}

\bibitem{berry} see for example M. Berry, {\it Principles of Cosmology
and Gravitation}, Cambridge University Press (1976)

\bibitem{gef}
G. Esposito-Far\`ese \& D. Polarski, {\it Scalar tensor gravity in an
accelerated Universe,} gr-qc/0009034, Phys. Rev. D63 (2001) 063504

\bibitem{r} A. G. Riess et al. Astroph. J. 607 (2004) 665

\bibitem{jj}
S. Jha , P. M. Garnavich, R. P. Kirshner et al., {\it The Type Ia
Supernova 1998 BU in M96
and the Hubble Constant}, ApJS 125 (1999) 73
\bibitem{ss}
A. Saha, A. Sandage, G. A. Tamman et al., {\it Cepheid calibration of
the peak brightness of
Type Ia supernovae XI. Sn 1998ap in NGC 3982}, ApJ 562 (2001) 314
\bibitem{gs}
B. K. Gibson \& P. B. Stetson, {\it Supernova 1991T and the value of the
Hubble Constant},
ApJ 547 (2001) L103
\bibitem{kr}
L. M. Krauss, {\it Space, time and matter: Cosmological Parameters 2001}, in
{\it Identification of Dark Matter} (2001)
\bibitem{dd}
J. Raux, {\it Photom\'etrie diff\'erentielle de supernovae de
type Ia lointaines $(0.5<z<1.2)$
mesur\'ees avec le t\'elescope spatial Hubble et estimation des
param\`etres cosmologiques}, PhD Universit\'e Paris 11 (2003)
\bibitem{ff}
F. Henry-Couannier et al.,
{\it Negative Energies and a Constantly Accelerating Flat Universe}
to be published



\end{thebibliography}
\end{document}